\documentclass[aps,prl,nofootinbib,showpacs,twocolumn]{revtex4-1}%
\usepackage[T1]{fontenc}
\usepackage[english]{babel}
\usepackage{graphicx}
\usepackage{epsfig}
\usepackage{epstopdf}
\usepackage{amsmath}
\usepackage{amssymb}
\usepackage{times}
\usepackage{setspace}
\usepackage{verbatim}
\usepackage{commath}
\usepackage{color}

\begin{document}
\title{Polyelectrolyte solution under spatial and dielectric confinement}
\author{Debarshee Bagchi}
\email[E-mail address:]{debarshee.bagchi@northwestern.edu}
\author{Trung Dac Nguyen}
\email[E-mail address:]{trung.nguyen@northwestern.edu}
\author{ Monica Olvera de la Cruz}
\email[E-mail address:]{m-olvera@northwestern.edu}
\affiliation{Department of Materials Science and Engineering, Northwestern University, Evanston, IL 60208}
\date{\today}

%===================================================================================================================================================================================
%===================================================================================================================================================================================

\begin{abstract}
Polyelectrolytes under confinement are crucial for energy storage and for understanding biomolecular functions. Using molecular dynamics simulations, we analyze a polyelectrolyte 
solution confined between two oppositely charged planar dielectric surfaces and include surface polarization effects due to dielectric mismatch at the two electrodes. Although the 
effect of polarization on the charge distribution seems minor, we find that surface polarization enhances energy storage and also leads to the emergence of negative differential 
capacitance in confined polyelectrolyte solutions.
\end{abstract}
\pacs{}
\maketitle

%===================================================================================================================================================================================
%===================================================================================================================================================================================
Polyelectrolytes exhibit a plethora of physical properties \cite{raspaud1998precipitation,netz2003neutral, dobrynin2005theory, buyukdagli2017like} of great interest in physical 
and life science, and technology. Proteins and nucleic acids are heterogeneous polyelectrolytes critical to biological function and biotechnology \cite{levin2005strange, 
luo2000synthetic, cakmak2004synthesis} while synthetic polyelectrolytes are components of modern technologies \cite{huang2015self,gonzalez2016review, vangari2012supercapacitors, 
simon2010materials}. Various functions of polyelectrolytes in biological settings and in technological applications take place in confinement. Supercapacitors, also referred to as 
electrical double-layer capacitors, for example, have attracted attention because of their long life cycle, fast charging and discharging, good charge density as well as 
power density, reliable performance over a large temperature range, and low maintenance requirements. In a supercapacitor, usually ionic liquids or aqueous electrolyte solutions 
are confined between two carbon-based electrodes like graphene \cite{liu2010graphene, yang2017graphene}. In confinement, both simple and molecular electrolytes display 
intriguing phenomena, such as charge inversion and overcharging \cite{jimenez2004model}, breakdown of local charge neutrality \cite{colla2016charge}, enhanced repulsions 
\cite{perez2017scaling} or attractions \cite{zwanikken2013tunable}, enhanced mobility \cite{antila2018dielectric, li2016ionic}, and non-monotonic electrophoretic mobility 
\cite{hickey2013electrophoretic}.

Electrolytes and polyelectrolytes in contact with one or two oppositely charged surfaces have been extensively studied \cite{dobrynin2005theory, qiao2011atomistic, jing2015ionic,
colla2016charge, guerrero2014polarization, girotto2018lattice, dos2016simulationsa, dos2016simulationsb, dos2015electrolytes, antila2018dielectric, perez2017scaling, 
messina2004effect}. In reality, the dielectric constant of the charged surfaces can be very different from that of the electrolyte or the polyelectrolyte. However, the dielectric 
discontinuities at the electrodes require taking into account the effects of surface polarization, which is a computationally non-trivial task \cite{jing2015ionic, 
dos2015electrolytes, girotto2018lattice}. For this reason, most studies do not include surface polarization or chose to simplify the model by taking into account only one 
dielectric discontinuity \cite{messina2004effect, antila2018dielectric}. Here, along with the geometrical confinement due to the two charged impenetrable surfaces, we study the 
effect of dielectric confinement on a model polyelectrolyte system. We consider a salt-free aqueous polyelectrolyte solution confined between two oppositely charged planar 
electrodes with explicit counterions in an implicit solvent. The adsorption of the polyelectrolyte is investigated here in terms of properties of the charge accumulations near the 
two electrodes, commonly referred to as the {\it electric double-layers}.

We find that, unlike equivalent bulk polyelectrolytes with monovalent counterions or in symmetrical confined molecular electrolytes, confined polyelectrolyte solutions exhibit 
features such as charge amplification and charge inversion. Moreover, the energy storage in the double-layer for a confined polyelectrolyte solution is found to be larger than that 
of a monovalent symmetric electrolyte solution under confinement. Furthermore, in presence of surface polarization effects, the energy stored is enhanced, which has possible 
technological implications. A crucial quantity to characterize double-layer properties is the differential capacitance that relates the change in charge storage to a small change 
in the voltage across the double-layer. We show that, keeping every other parameter the same, when the effects of surface polarization are included one obtains a negative 
differential capacitance for the double-layer formed by the polyelectrolyte molecules. Thus, although surface polarizability effects on polyelectrolyte adsorption seem to be small 
as has also been seen in some recent works \cite{messina2004effect, messina2006erratum, dos2016simulationsb}, they play an important role in the energy storage and differential 
capacitance of a confined polyelectrolyte solution.

%\paragraph{Model and simulation method:}

%
\begin{figure}[htb]
\centering
{\includegraphics[width=5cm]{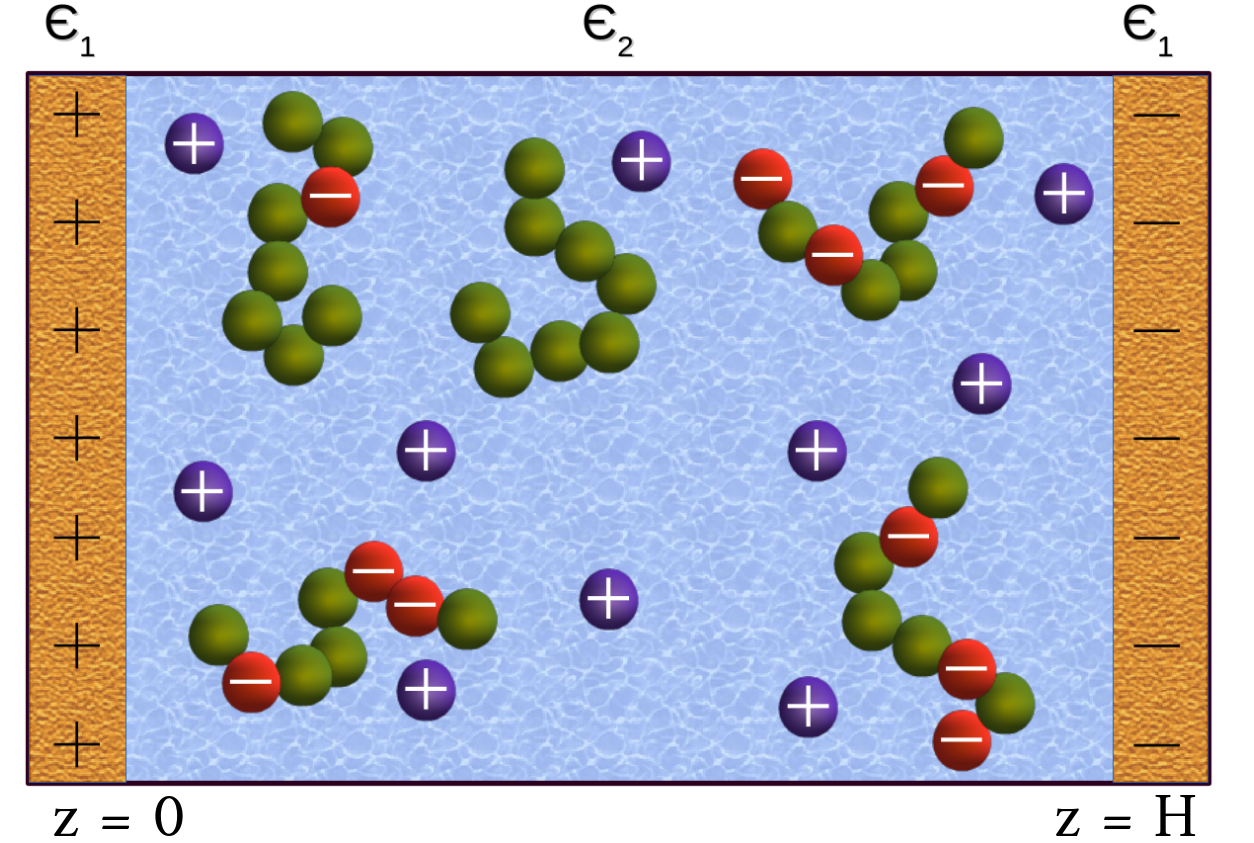}}
\caption{Schematic diagram of the polyelectrolyte model confined between two dielectric electrodes.}
\label{fig:model}
\end{figure}
A schematic representation of the model system is depicted in Fig. \ref{fig:model}. In the course grained molecular dynamics (CGMD) simulations, the polyelectrolyte is represented 
by a linear bead-spring chain with randomly distributed charged beads \cite{stevens1993structure}. The number of charged beads on the polyelectrolyte chains is controlled by the 
charge fraction $f_q = n/N$, where $n$ is the total number of charged monomer beads, each with one electron charge $-e$, and $N$ is the total number of monomers. For overall
electroneutrality we have $n$ positive counterion beads with $+e$ charge. 
The simulation box dimensions are $L_x = L_y = L = 30 \sigma$ and $Lz = H = 100 \sigma$, where our length unit $\sigma$ is the diameter of all the beads used in the simulation (we 
set $\sigma = 1$ which corresponds to $\sigma = 0.3$ nm in real units).
The two planar electrodes are also composed of beads and are oppositely charged. Each electrode has a charge density $\Sigma$ and are located at  $z=0$ (positive electrode) and 
$z=H$ (negative electrode). The electrodes are considered to be composed of a material of low dielectric constant $\epsilon_1 = 2$. The solvent is a continuum background with 
uniform dielectric constant $\epsilon_2 = 80$ and is a poor solvent for the partially charged polyelectrolyte chains.
The $N_m = 40$ monomers on each of the $N_c = 60$ polyelectrolyte chains interact via a truncated-shifted Lennard Jones (LJ) potential
$V^{LJ}_{ij}(r) = 4\varepsilon \left[(\sigma/r)^{12} - (\sigma/r)^6 \right]$,
where $\varepsilon = k_B T$ is the energy scale which we set to unity in reduced units, cutoff $r_c = 2.5\sigma$ such that $V_{LJ}(r > r_c) = 0$, and $T$ is the temperature. All 
other short-range interactions are purely repulsive LJ potential (Weeks-Chandler-Andersen) with $r_c = 2^{1/6} \sigma.$ Two charged beads interact with each other via the 
long-range Coulomb potential
$V^{q}_{ij}(r) = k_B T l_B \frac{q_i q_j}{r}$,
where $q_i, q_j$ are the valencies of the charged beads and $l_B$ is the Bjerrum length
$l_B = \frac{e^2}{\epsilon_0 \epsilon_r k_B T}$ (all the symbols have their usual meaning); we set $l_B = 0.7$ nm for water.
Coulomb interactions among charged monomers and counterions are computed using $\epsilon_r = \epsilon_2$ and with the surface beads using $\epsilon_r = \frac 12 (\epsilon_1 + 
\epsilon_2)$ \cite{tyagi2010iterative} (see Fig. \ref{fig:model}),
and are computed using the particle-particle particle-mesh (PPPM) method with an accuracy of $10^{-4}$ and slab corrections \cite{frenkel2001understanding}.
The consecutive beads in the polyelectrolyte chain are connected by finite extensible nonlinear elastic (FENE) bonds $V^b(r) = - \frac 12 k R_0^2 \ln \left(1- \frac{r^2}{R_0^2} 
\right)$, with maximum bond extension $R_0 = 1.5 \sigma$ and spring constant $k = 30\varepsilon/\sigma^2$.

\begin{figure}[htb]
%\centering
\hskip-0.25cm
{\includegraphics[width=4.25cm]{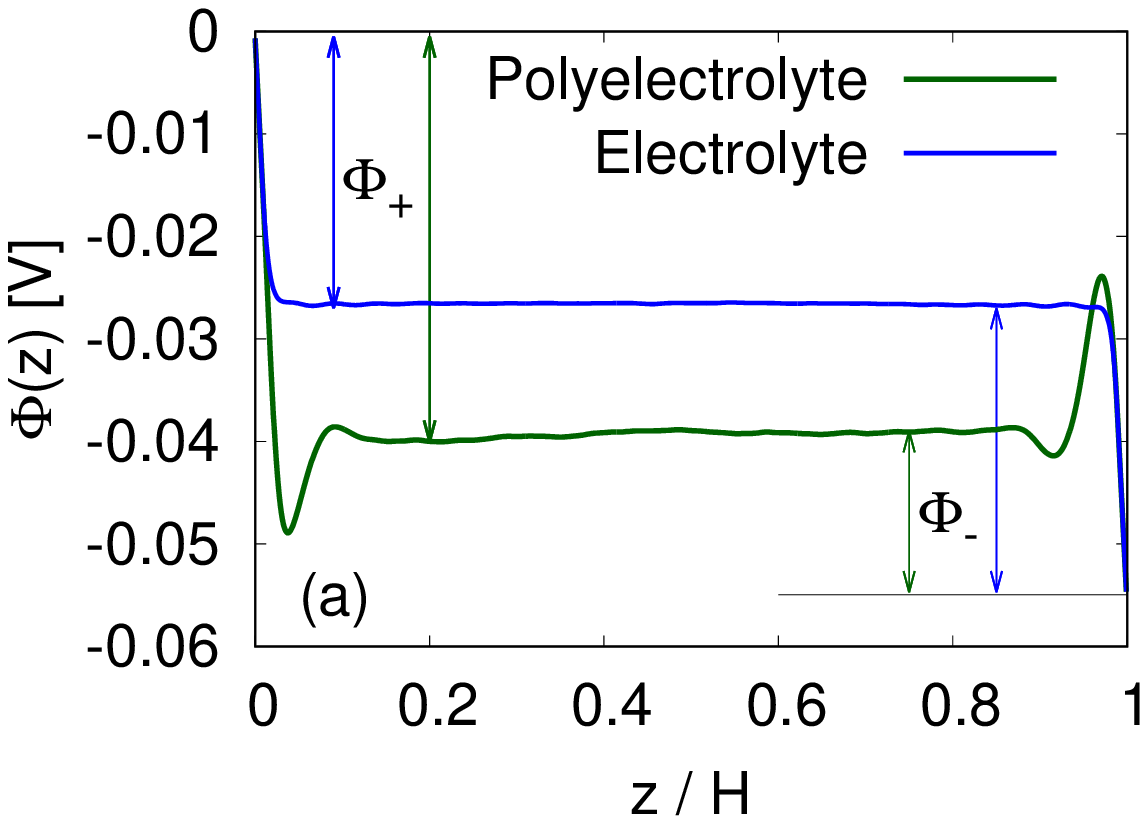}} \hskip-0.15cm
{\includegraphics[width=4.25cm]{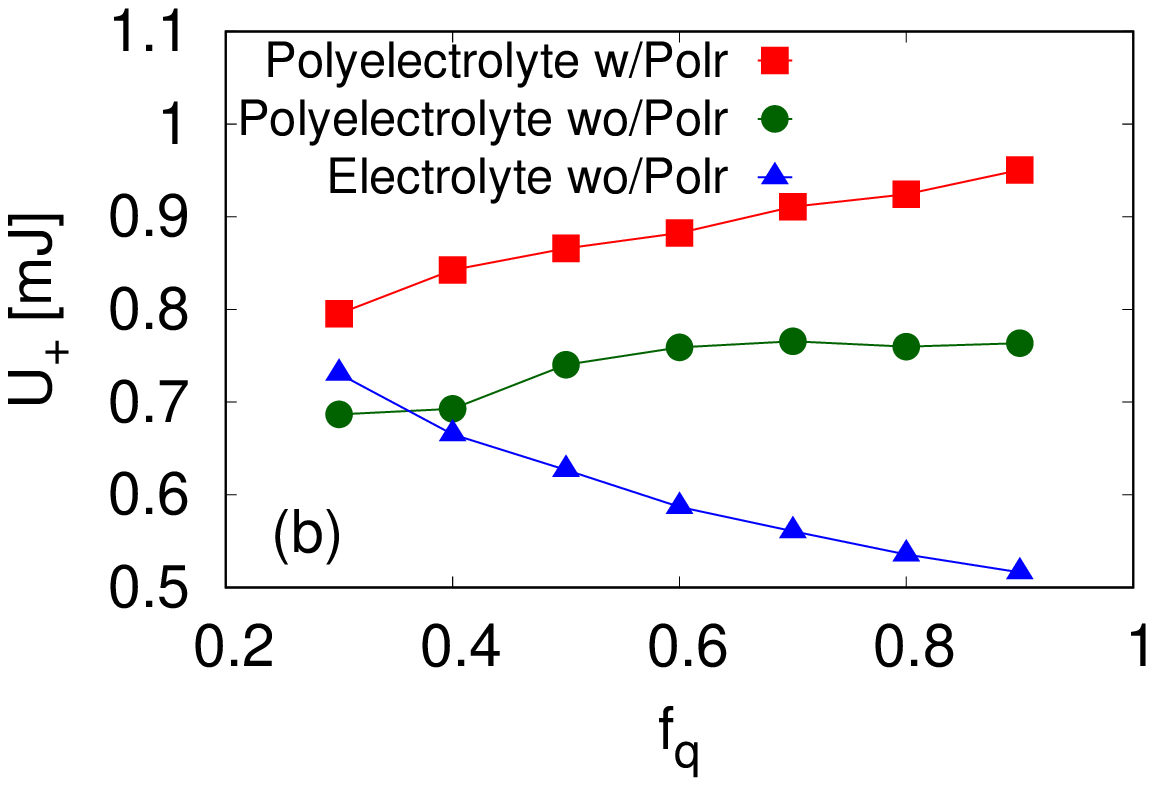}}
{\includegraphics[width=4.25cm]{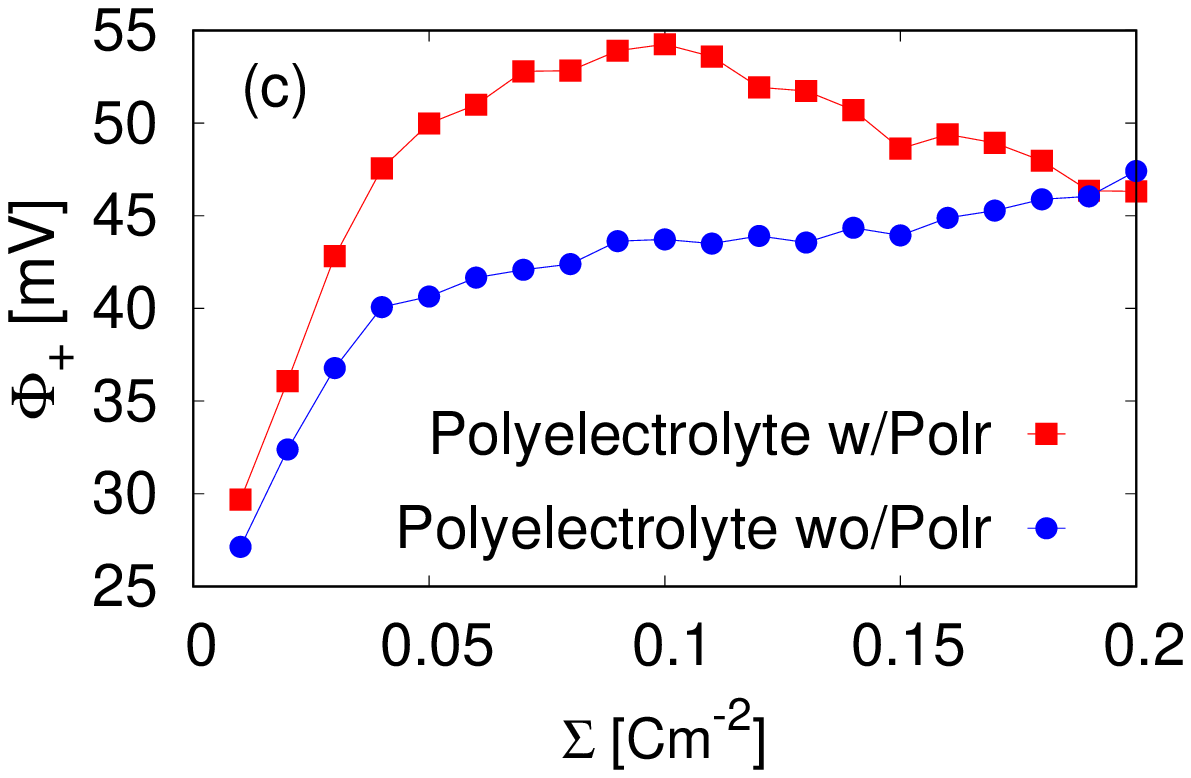}} \hskip-0.2cm
{\includegraphics[width=4.25cm]{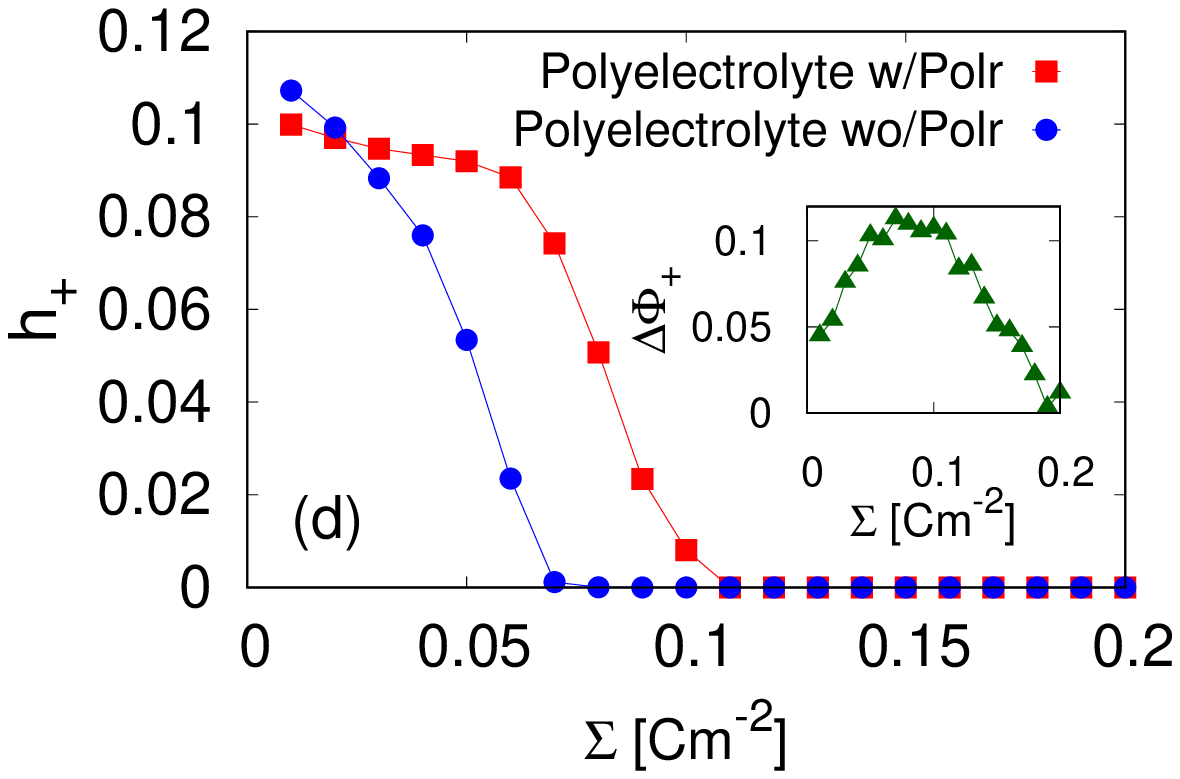}}
\caption{(a) Typical Poisson potential $\Phi(z)$ profiles of polyelectrolyte and electrolyte for charge fraction $f_q = 0.90$. (b) The double-layer energy $U_+$ with charged 
fraction $f_q$ for the electrolyte (filled triangles) and polyelectrolyte (filled circles), both without polarization, and polyelectrolyte with polarization (filled squares).
The variation of (c) the EDL potential $\Phi_+$ and (d) the height of the charge amplification peak for different values of surface charge density $\Sigma$ with (filled 
squares) and without (filled circles) polarization effects. The inset in (d) shows the variation of $\Delta \Phi_+$ (see main text) with surface charge density $\Sigma$.}
\label{fig:5abcd}
\end{figure}

We integrate the equations of motion using a symplectic Verlet-velocity algorithm in the canonical ensemble with Langevin thermostat \cite{frenkel2001understanding} using a 
reduced time-step $\Delta t = 0.005$. Starting from random initial configurations for the monomer and the counterion beads, their equations of motion are evolved for $\sim10^5$ MD 
time-steps and thereafter average quantities are computed 
for another $\sim10^6$ time-steps. For most of the results (except Fig. \ref{fig:5abcd}c,d) the electrode charge density is set to $\Sigma = 0.04 ~Cm^{-2}$. The effects of 
polarizability 
due to dielectric mismatch is taken into account by employing the ICC* (Induced Charge Computation) method \cite{tyagi2010iterative}; for details of implementation of this 
algorithm in LAMMPS see Ref. \cite{nguyen2019incorporating}. Briefly, in order to calculate the bound charges due to the dielectric mismatch, we discretize the two interfaces 
into square grids of area $\sim 0.82 \sigma^2$ and use an iterative scheme to obtain the bound charges from the free charges. The computationally expensive 
calculation of the polarization charges becomes manageable using the ICC* algorithm and yields verifiably accurate results \cite{nguyen2019incorporating}.

\begin{figure}[htb]
\centering
%\hskip-0.25cm
{\includegraphics[width=5.75cm]{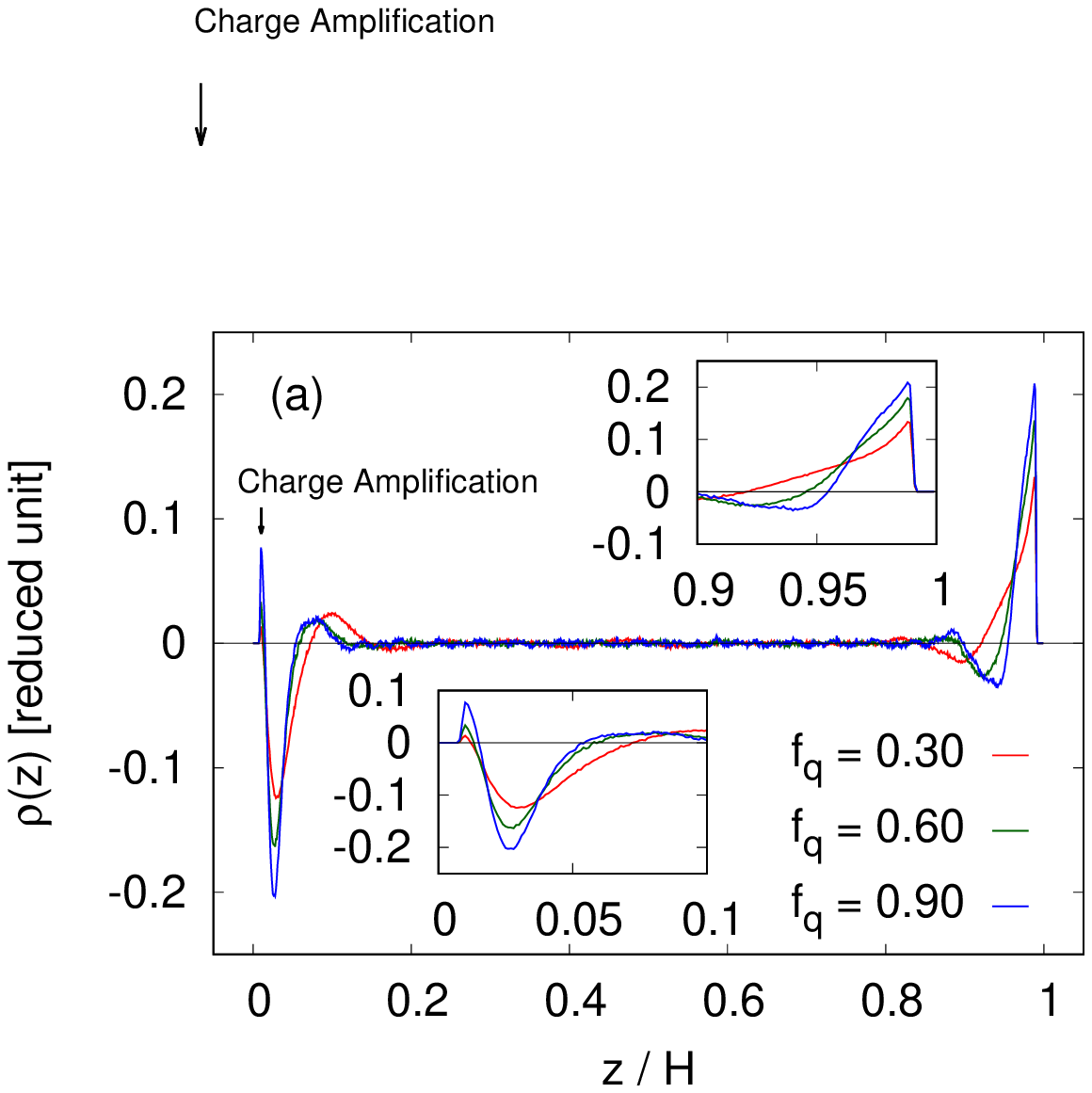}} %\hskip-0.15cm
{\includegraphics[width=5.75cm]{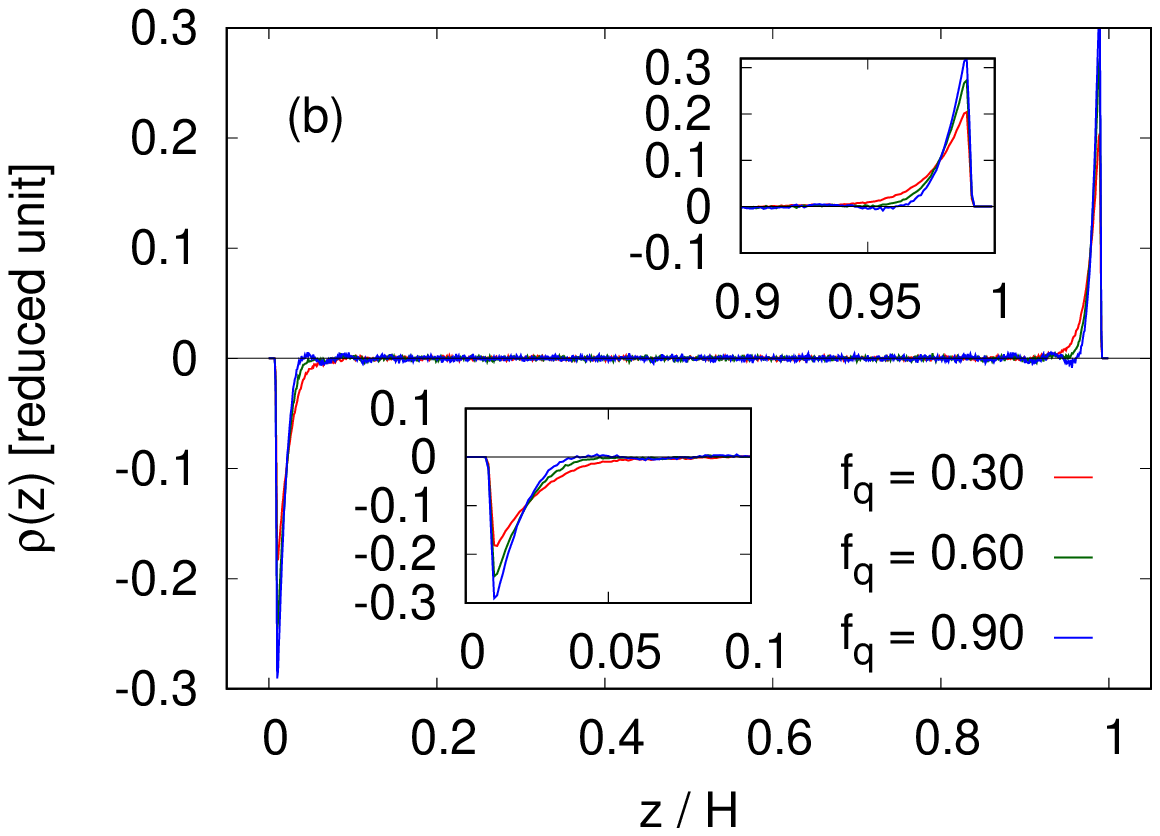}}
{\includegraphics[width=4.25cm]{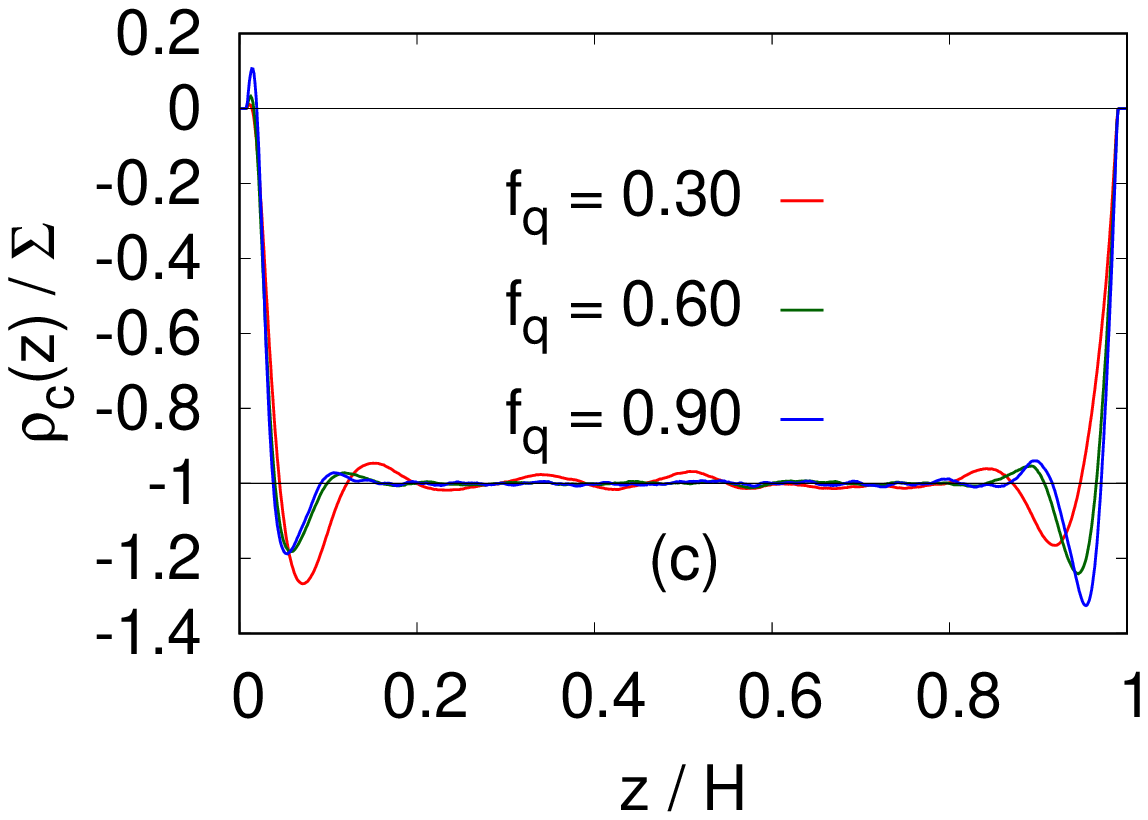}} \hskip-0.15cm
{\includegraphics[width=4.25cm]{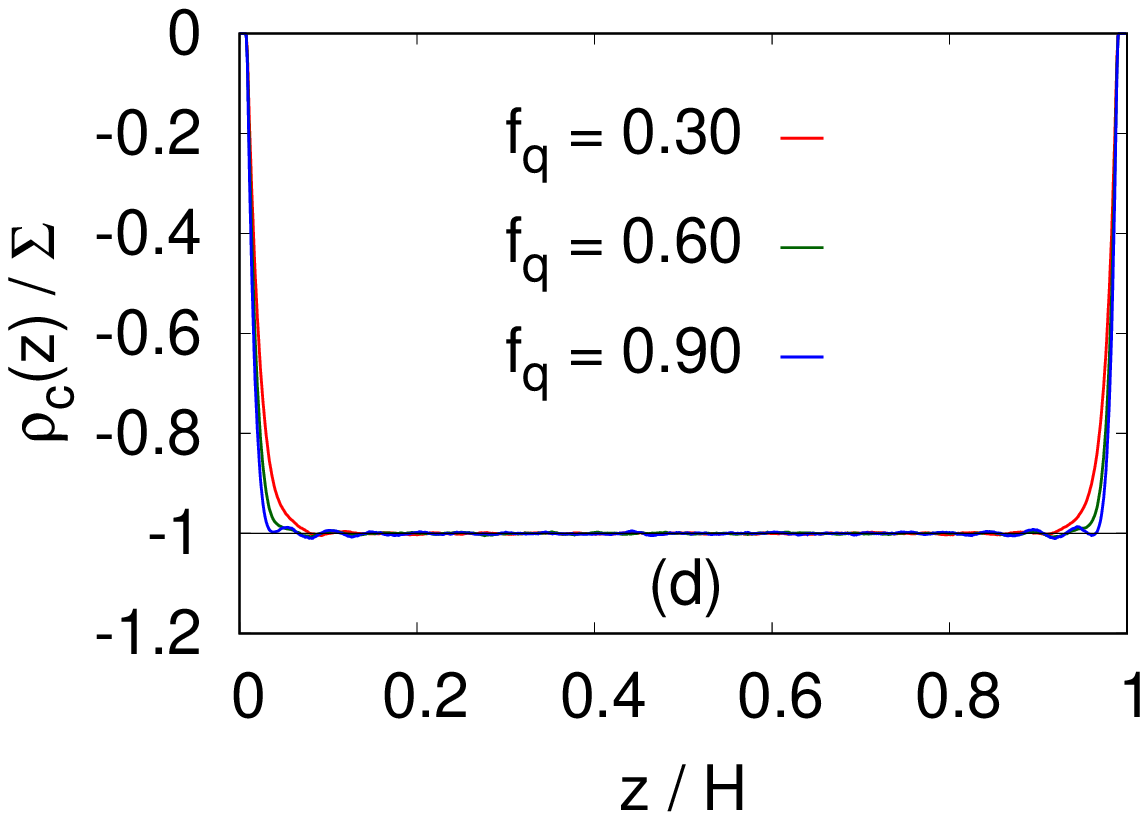}}
\caption{Net charge density profile $\rho(z)$ for different values of the charge fraction $f_q$ for (a) polyelectrolyte solution and (b) electrolyte solution, with all parameters 
being the same. The insets in both the plots magnify the density profile near the two electrodes. Cumulative charge density $\rho_c(z)$ scaled by the interface charge density 
$\Sigma$ for different values of the charge fraction $f_q$ for (c) polyelectrolyte solution and (d) electrolyte solution, computed from the data in (a) and (b) respectively.}
\label{fig:2abcd}
\end{figure}

%\paragraph{Charge density without surface polarization:}
In order to benchmark the confined polyelectrolyte system, we first run the simulations without surface polarization effect. The time-averaged net charge 
density profile along the direction of confinement $z$ is defined as $\rho(z) = \rho_+(z) - \rho_-(z) $, where $\rho_{+}$ and $\rho_{-}$ correspond to the charge densities of the 
positively charged counterions and the negatively charged monomers respectively. In Fig. \ref{fig:2abcd}a, $\rho(z)$ is depicted for three different values of the charge fraction 
$f_q$. We find that the negatively charged polyelectrolyte chains accumulate near the positive electrode at $z=0$, whereas the positively charged counterions are seen to
accumulate near both the electrodes. This accumulation of counterions near a like-charged electrode is referred to as {\it charge amplification} (also referred to as {\it 
overcharging}) \cite{jimenez2004model,guerrero2010overcharging,guerrero2010effects} and becomes more pronounced for higher $f_q$ values as can be clearly seen in the inset. This 
phenomenon arises because the gain in entropy of the counterions is larger than the repulsion it experiences due to the like-charged electrode. However, charge amplification 
is observed only for low electrode charge densities $\Sigma$; for larger $\Sigma$ values the repulsion between the electrode and the counterions becomes large and charge 
amplification disappears.
In Fig. \ref{fig:2abcd}b we show $\rho(z)$ for an equivalent electrolyte solution, which is identical to the polyelectrolyte solution except that all bond potentials are set to 
zero. The net charge density profile does not show charge amplification and this, as we show below, affects the energy storage of the two systems considerably.

Fig. \ref{fig:2abcd}c shows that the cumulative charge density $\rho_c(z) = \int_0^z \rho(s) ds$ for the polyelectrolyte solution near both the electrodes is larger in magnitude 
than the electrode charge density $\Sigma$. This implies that each electrode attracts more charges of the opposite sign than what is necessary to neutralize its charge, which is a 
phenomenon known as {\it charge inversion} \cite{messina2002overcharging,guerrero2010effects}. Note that, here charge amplification (the positive part of the curve in Fig. 
\ref{fig:2abcd}c) is observed only in the double-layer near the positive electrode but charge inversion is observed near both the electrodes. Thus, the polyelectrolyte solution 
exhibits charge inversion even if there is no charge amplification, while all these effects are absent in the electrolyte system shown in Fig. \ref{fig:2abcd}d.
%
% \begin{figure}[htb]
% %\centering
% \hskip-0.25cm
% {\includegraphics[width=4.25cm]{overcharge.eps}} \hskip-0.15cm
% {\includegraphics[width=4.25cm]{overcharge_el.eps}}
% \caption{Cumulative charge density $\rho_c(z)$ scaled by the interface charge density $\Sigma$ for different values of the charge fraction $f_q$ for (a) polyelectrolyte solution 
% (b) electrolyte solution, computed from the data in Fig. \ref{fig:prof}a and b respectively.}
% \label{fig:overcharging}
% \end{figure}
%

\begin{figure}[htb]
%\centering
%\hskip-0.4cm
{\includegraphics[width=5.85cm]{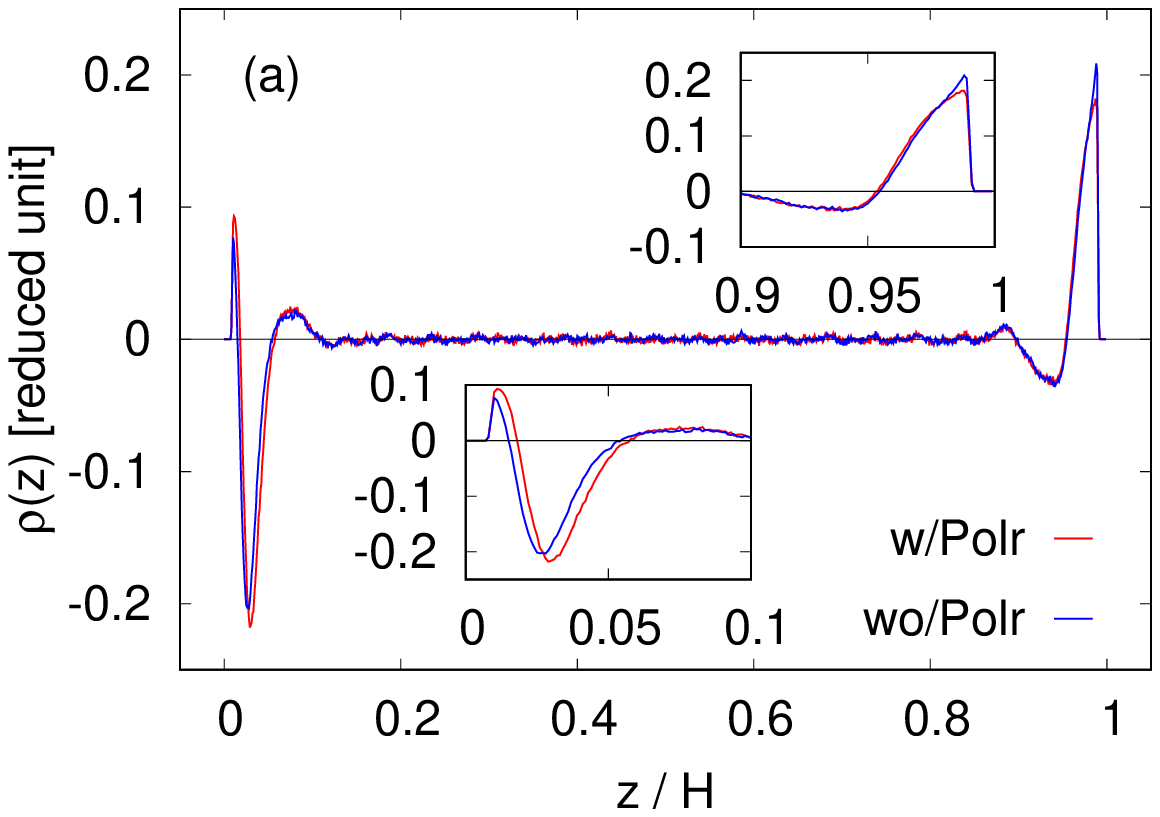}}
{\includegraphics[width=5.95cm]{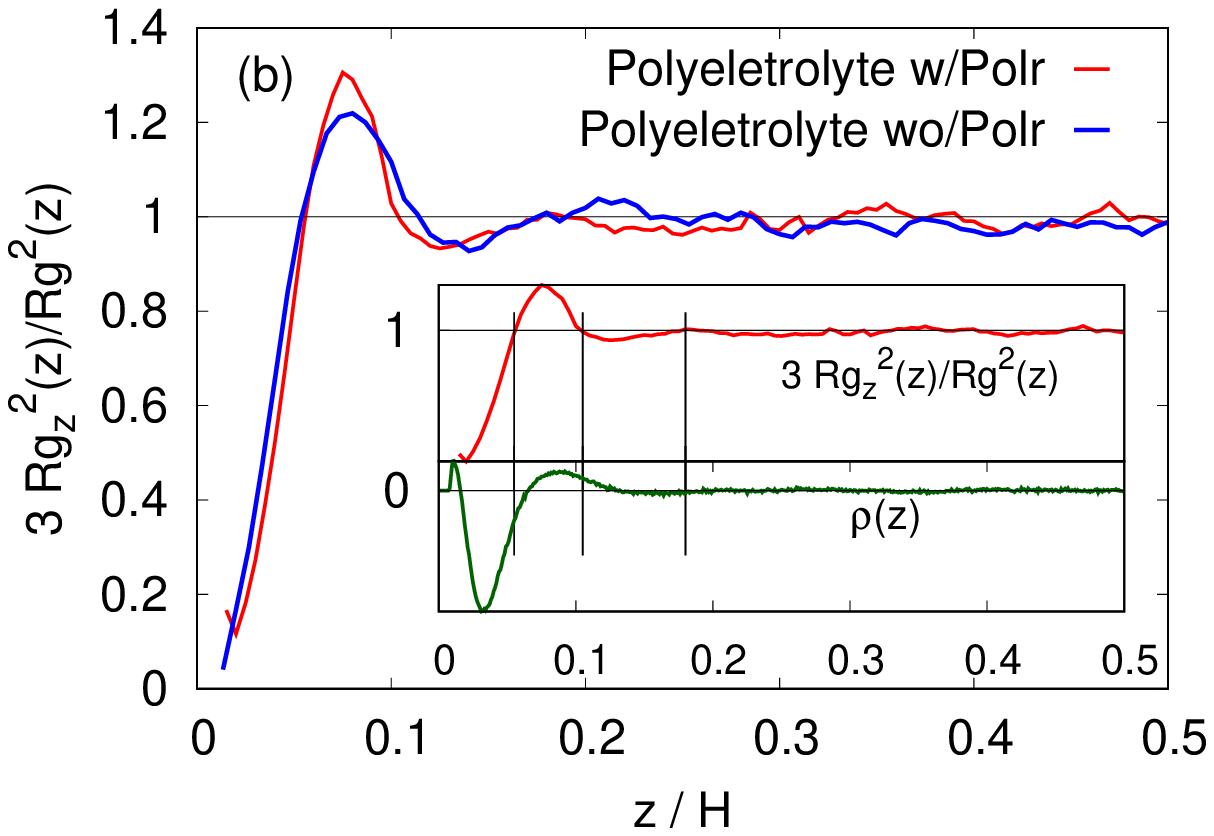}}
\caption{(a) Comparison of the charge density profile $\rho(z)$ of the polyelectrolyte solution for $f_q = 0.90$ with and without polarization effects. (b) The main figure shows 
$3R_{gz}^2(z)/R_g^2$ for the polyelectrolyte solution with and without polarization for $f_q = 0.40$. In the inset, $3R_{gz}^2(z)/R_g^2$ is shown against the net charge density 
$\rho(z)$ for the polyelectrolyte solution with polarization effects.}
\label{fig:polrnopolr}
\end{figure}

%\paragraph{Effect of surface polarizability:}
The effect of surface polarizability on the net charge density profile of the polyelectrolyte solution is shown in Fig. \ref{fig:polrnopolr}a. We find that the density profile of 
the double-layer is shifted away from the positive electrode (as can be seen in the inset of Fig. \ref{fig:polrnopolr}a); that is, the polyelectrolyte solution experiences more 
confinement when polarization effects are present. Therefore, the general effect of surface polarizability (when $\epsilon_1 < \epsilon_2$) is to add dielectric confinement on top 
of the physical confinement due to the impenetrable surfaces at the two ends of the simulation box. There are however subtle differences in how polarization affects different 
parts of the net charge density profile. A closer look at the two insets in Fig. \ref{fig:polrnopolr}a shows that the peak height of the double-layer formed near the negative 
electrode decreases in presence of polarization, whereas the peak heights near the positive electrode increase. Thus, in presence of surface polarizability, charge amplification 
becomes stronger. Although, at the level of the charge density profile, the effect of polarization looks quite small, we show in the following that polarization affects the energy 
storage and the capacitance of the polyelectrolyte solution significantly.
%

%\paragraph{Conformational properties:}
Interesting features in the conformation of the polyelectrolyte chains are observed in the double-layer formed near the positive electrode. We analyze this by computing the radius 
of gyration squared $R_g^2 = \frac 1{N_m} \sum_{i = 1}^{N_m} (\vec{r}_i - \vec{r}_{cm})^2$ and its components $R_{g\alpha}^2 = \frac 1{N_m} \sum_{i = 1}^{N_m} (\vec{r}_i^{\alpha} - 
\vec{r}_{cm}^{\alpha})^2$, where $\alpha \equiv x,y,z$. In Fig. \ref{fig:polrnopolr}b, $3R_{gz}^2(z)/R_g^2$ is shown with and without polarization effects. In the bulk, since 
$R_{gz}^2 = R_{g\perp}^2$ ($R_{g\perp}^2 \equiv R_{gx}^2, R_{gy}^2$), we obtain $3R_{gz}^2(z)/R_g^2 \approx 1$ (since $R_g^2 = R_{gz}^2 + 2R_{g\perp}^2$) and therefore the chains, 
on an average, have a spherically symmetric conformation. However, next to the left interface $3R_{gz}^2(z)/R_g^2 < 1$, which implies that $R_{gz}^2 < R_{g\perp}^2$. Thus, close to 
the left electrode, the polyelectrolyte chains are compressed along the $z-$ direction and assume an oblate-spheroid conformation. As one moves towards the bulk from the left 
electrode, $3R_{gz}^2(z)/R_g^2$ gradually increases and, interestingly enough, at one point it overshoots the bulk value of unity. In this overshoot region, the chains have a 
prolate-spheroid conformation and are more elongated along the $z$-direction compared to the perpendicular $x,y$-directions, that is, $R_{gz}^2 > R_{g\perp}^2$. This {\it 
$z$-compressed} oblate to a {\it $z$-elongated} prolate conformational change can be explained from the fact that the polyelectrolyte chains near the positive electrode try to 
minimize the electrostatic repulsion of the neighboring layers of polyelectrolyte chains. Consequently, since the confinement effects are stronger in the presence of polarization, 
this overshooting is more prominent, as seen in Fig. \ref{fig:polrnopolr}b (main figure).
%
% \begin{figure}[htb]
% %\centering
% %\hskip-0.4cm
% {\includegraphics[width=5.95cm]{Zvarn_Rgz_mul.eps}} \vskip-0.1cm
% %{\includegraphics[width=5.25cm]{./Figures/Cubatic.png}}
% \caption{}
% \label{fig:Zvarn}
% \end{figure}

To get a sense of where this {\it flipping} from {\it $z$-compressed} conformation to a {\it $z-$elongated} conformation happens, we plot $3R_{gz}^2(z)/R_g^2$ along with the net 
charge density profile $\rho(z)$ of the system in the inset of Fig. \ref{fig:polrnopolr}b. We find that the conformation flipping of the polyelectrolyte chains happens in a region 
which mostly has a net positive charge (i.e., excess counterions). Also, a closer look at the data in Fig. \ref{fig:polrnopolr}b reveals that this oblate-prolate flipping happens 
not once but at least twice, before $3R_{gz}^2(z)/R_g^2$ becomes equal to unity in the bulk of the system. Thus, the polyelectrolyte chains near the left interface arrange 
themselves in alternate layers of {\it z-compressed} and {\it $z-$elongated} conformations under confinement. This phenomenon is, in some sense, akin to that of the {\it cubatic} 
phase observed originally in simulation studies of hard cut spheres \cite{veerman1992phase}, although the scenario is much more complicated here.

%\paragraph{Energy storage:}
We measure the electrostatic energy storage in the confined polyelectrolyte solution by first computing the potential profile $\Phi(z)$ between the two electrodes. This is 
done by numerically integrating the Poisson's equation $\partial_z^2\Phi(z) = - \rho(z)/(\epsilon_0 \epsilon_r)$, where $\rho(z)$ is the net charge density profile. Typical 
potential profiles $\Phi(z)$ obtained from the simulations for the polyelectrolyte and the electrolyte solution are shown in Fig. \ref{fig:5abcd}a. We denote the potential drop 
between positive electrode and the bulk as $\Phi_+$ (and similarly for $\Phi_-$) which is the voltage drop across the electric double-layer (EDL) and referred to as the EDL 
potential. Note that, for the polyelectrolyte solution $\Phi_+ \neq \Phi_-$, unlike the electrolyte solution. Here, we will focus on the EDL potential $\Phi_+$ since this is the 
voltage drop across the double-layer formed near the positive electrode where the polyelectrolyte chains get absorbed. The electrostatic energy stored in this EDL (per unit 
electrode area) is given by $U_+ = \frac 12 \Phi_+  Q_+ / L^2$, where $Q_+$ is the total charge in the EDL which is obtained from the simulations. The energy storage $U_+$ with 
charge fraction $f_q$ is shown in Fig. \ref{fig:5abcd}b. It is found that the energy storage in the polyelectrolyte EDL is higher than the energy stored in the electrolyte EDL. 
Moreover, we see that energy stored for the polyelectrolyte solution is further enhanced when one takes into account the effects of surface polarizability.

%\paragraph{Differential capacitance:}
Next, in order to find the differential capacitance $C_d$, we compute first the double-layer potential $\Phi_+$ for different values of the surface charge density $\Sigma$ (Fig. 
\ref{fig:5abcd}c). The double-layer potential $\Phi_+$ for the polyelectrolyte in presence of polarization effects is considerably higher than $\Phi_+$ in absence of polarization. 
Moreover, for the polyelectrolyte {\it without} polarizability the $\Phi_+$ {\it vs} $\Sigma$ curve is monotonically increasing. (For a monovalent symmetric electrolyte solution 
without polarization the increase is found to be linear with a roughly constant slope for the same range of $\Sigma$ values). However, for the polyelectrolyte {\it with} 
polarizability effects $\Phi_+$ {\it vs} $\Sigma$ is seen to be non-monotonic. The differential capacitance $C_d$ is defined as the inverse slope of the $\Phi_+$ {\it vs} $\Sigma$ 
curves, $C_d = \left(d\Phi_+/d\Sigma \right)^{-1}$, and thus, it implies that, in presence of polarization effects one obtains a {\it negative differential capacitance}, $C_d < 0$. 
As is well known, equilibrium thermodynamics dictates that the differential capacitance should always be strictly positive because of stability reasons. However, the possibility of 
a negative differential capacitance under various conditions has been actively investigated (see, for example, \cite{partenskii1996question, partenskii2008limitations, 
partenskii2009squishy}). Partenskii and coworkers, suggested that the anomalous $C_d < 0$ can emerge in a uniform {\it charge density controlled} system like ours and this implies 
that there could be interfacial instabilities and charging induced surface phase transitions in a constant {\it potential controlled} systems \cite{partenskii1996question}. Such 
anomalous behavior of the differential capacitance in a uniform surface charge density setup has also been suggested to explain some experimental results 
\cite{partenskii2008limitations} recently. Thus, these calculations show that, inclusion of polarization effects not only changes the result in a quantitative way but also 
significantly changes the qualitative nature of the result.
%
% \begin{figure}[htb]
% %\centering
% \hskip-0.25cm
% {\includegraphics[width=4.25cm]{QV.eps}} \hskip-0.2cm
% {\includegraphics[width=4.25cm]{peak0mul.eps}}
% \caption{The variation of (a) the EDL potential $\Phi_+$ and (b) the height of the charge amplification peak for different values of surface charge density $\Sigma$ with (filled 
% squares) and without (filled circles) polarization effects. The inset in (b) shows $\Delta \Phi_+$ (see main text) for different values of surface charge density $\Sigma$.}
% \label{fig:diffcapa}
% \end{figure}

%\paragraph{Discussion:} 
The disparity in the behavior of the potential $\Phi_+$ in presence and absence of polarizability (Fig. \ref{fig:5abcd}c) can be intuitively explained by looking at charge 
amplification (Fig. \ref{fig:5abcd}d). If we denote by $h_+$ the height of the charge amplification peak (the peak next to the positive electrode in Fig. \ref{fig:2abcd}a), we 
find from Fig. \ref{fig:5abcd}d that charge amplification in the polyelectrolyte {\it without} polarizability effects disappears at $\Sigma = 0.07 ~Cm^{-2}$, whereas the same 
lingers on till $\Sigma = 0.11~ Cm^{-2}$ in the presence of polarization effects. This stronger overcharging of the interface leads to a larger value of $\Phi_+$ in the presence of 
polarization. Thus, although the actual microscopic picture for the emergence of negative differential capacitance seems to be a complicated interplay of charge amplification, 
electrode charge density $\Sigma$, complex changes in polyelectrolyte conformations, etc., one of the reasons seems to be the stronger charge amplification in the presence of 
surface polarizability effect \cite{guerrero2014polarization}.

Another crucial point to note in Fig. \ref{fig:5abcd}c is regarding the importance of polarizability and its dependence on electrode charge density $\Sigma$. Although the general 
belief is that polarizability effects become increasingly unimportant as $\Sigma$ increases (see \cite{hatlo2008role} and references therein), Fig. \ref{fig:5abcd}c 
demonstrates that this is not true for the electrostatic potential. If we define $\Delta \Phi_+ = \frac{|\Phi^{'}_+ - \Phi^{0}_+|}{\Phi^{'}_+ + \Phi^{0}_+}$ as an indicator of the 
importance of polarization effects, where $\Phi_+^{'}$ and $\Phi_+^{0}$ are the $\Phi_+$ potential with and without polarizability, we find that $\Delta \Phi_+$ {\it vs} $\Sigma$ 
has a maximum close to $\Sigma = 0.07 ~Cm^{-2}$, as shown in the inset of Fig. \ref{fig:5abcd}d. Thus, the effect of polarizability on the electrostatic potential is found to be 
most prominent for $\Sigma \approx 0.07 ~Cm^{-2}$ and small on either side of this $\Sigma$ value.

An intuitive way to understand why there is a large effect of polarizability on the energy storage (Fig. \ref{fig:5abcd}b) and differential capacitance (Fig. \ref{fig:5abcd}c) 
but only a nominal effect on the charge density profiles (Fig. \ref{fig:polrnopolr}) is by noting that both energy and differential capacitance depend on the potential $\Phi$, 
which is a double integral of the charge density profile $\rho$ over the entire confinement, $\Phi(z) \sim \int dz' \int dz'' \rho(z'')$. Thus, the small differences in 
$\rho(z)$ with and without polarization add up, resulting in large changes in the energy storage and the differential capacitance. 

To summarize, motivated by the design of a super-capacitor, a polyelectrolyte solution confined between two oppositely charged planar dielectric surfaces is studied here via 
efficient coarse-grained molecular dynamics simulations that include the dielectric discontinuities at the two electrodes. At the level of charge density profiles, the effect due 
to surface polarization is small. However, surface polarization changes the energy storage and differential capacitance of the confined polyelectrolyte solution significantly. The 
electrostatic energy stored in the polyelectrolyte double-layer is found to  be higher than that of an electrolyte solution, and it is further enhanced when surface polarization 
effects are taken into account. We also find that the confined polyelectrolyte solution exhibits negative differential capacitance in the presence of surface polarization effects.

{\it Acknowledgement:}
This work has been funded by NSF DMR Award No. $1611076$. We also thank the computational support of Sherman Fairchild Foundation. DB gratefully acknowledges many fruitful 
discussions with F. J.-\'{A}ngeles and A. Ehlen for careful reading of the manuscript.

 \newcommand{\noop}[1]{}


\begin{thebibliography}{41}%
\makeatletter
\providecommand \@ifxundefined [1]{%
 \@ifx{#1\undefined}
}%
\providecommand \@ifnum [1]{%
 \ifnum #1\expandafter \@firstoftwo
 \else \expandafter \@secondoftwo
 \fi
}%
\providecommand \@ifx [1]{%
 \ifx #1\expandafter \@firstoftwo
 \else \expandafter \@secondoftwo
 \fi
}%
\providecommand \natexlab [1]{#1}%
\providecommand \enquote  [1]{``#1''}%
\providecommand \bibnamefont  [1]{#1}%
\providecommand \bibfnamefont [1]{#1}%
\providecommand \citenamefont [1]{#1}%
\providecommand \href@noop [0]{\@secondoftwo}%
\providecommand \href [0]{\begingroup \@sanitize@url \@href}%
\providecommand \@href[1]{\@@startlink{#1}\@@href}%
\providecommand \@@href[1]{\endgroup#1\@@endlink}%
\providecommand \@sanitize@url [0]{\catcode `\\12\catcode `\$12\catcode
  `\&12\catcode `\#12\catcode `\^12\catcode `\_12\catcode `\%12\relax}%
\providecommand \@@startlink[1]{}%
\providecommand \@@endlink[0]{}%
\providecommand \url  [0]{\begingroup\@sanitize@url \@url }%
\providecommand \@url [1]{\endgroup\@href {#1}{\urlprefix }}%
\providecommand \urlprefix  [0]{URL }%
\providecommand \Eprint [0]{\href }%
\providecommand \doibase [0]{http://dx.doi.org/}%
\providecommand \selectlanguage [0]{\@gobble}%
\providecommand \bibinfo  [0]{\@secondoftwo}%
\providecommand \bibfield  [0]{\@secondoftwo}%
\providecommand \translation [1]{[#1]}%
\providecommand \BibitemOpen [0]{}%
\providecommand \bibitemStop [0]{}%
\providecommand \bibitemNoStop [0]{.\EOS\space}%
\providecommand \EOS [0]{\spacefactor3000\relax}%
\providecommand \BibitemShut  [1]{\csname bibitem#1\endcsname}%
\let\auto@bib@innerbib\@empty
%</preamble>
\bibitem [{\citenamefont {Raspaud}\ \emph {et~al.}(1998)\citenamefont
  {Raspaud}, \citenamefont {Olvera de~la Cruz}, \citenamefont {Sikorav},\ and\
  \citenamefont {Livolant}}]{raspaud1998precipitation}%
  \BibitemOpen
  \bibfield  {author} {\bibinfo {author} {\bibfnamefont {E.}~\bibnamefont
  {Raspaud}}, \bibinfo {author} {\bibfnamefont {M.}~\bibnamefont {Olvera de~la
  Cruz}}, \bibinfo {author} {\bibfnamefont {J.-L.}\ \bibnamefont {Sikorav}}, \
  and\ \bibinfo {author} {\bibfnamefont {F.}~\bibnamefont {Livolant}},\
  }\href@noop {} {\bibfield  {journal} {\bibinfo  {journal} {Biophysical
  Journal}\ }\textbf {\bibinfo {volume} {74}},\ \bibinfo {pages} {381}
  (\bibinfo {year} {1998})}\BibitemShut {NoStop}%
\bibitem [{\citenamefont {Netz}\ and\ \citenamefont
  {Andelman}(2003)}]{netz2003neutral}%
  \BibitemOpen
  \bibfield  {author} {\bibinfo {author} {\bibfnamefont {R.~R.}\ \bibnamefont
  {Netz}}\ and\ \bibinfo {author} {\bibfnamefont {D.}~\bibnamefont
  {Andelman}},\ }\href@noop {} {\bibfield  {journal} {\bibinfo  {journal}
  {Physics Reports}\ }\textbf {\bibinfo {volume} {380}},\ \bibinfo {pages} {1}
  (\bibinfo {year} {2003})}\BibitemShut {NoStop}%
\bibitem [{\citenamefont {Dobrynin}\ and\ \citenamefont
  {Rubinstein}(2005)}]{dobrynin2005theory}%
  \BibitemOpen
  \bibfield  {author} {\bibinfo {author} {\bibfnamefont {A.~V.}\ \bibnamefont
  {Dobrynin}}\ and\ \bibinfo {author} {\bibfnamefont {M.}~\bibnamefont
  {Rubinstein}},\ }\href@noop {} {\bibfield  {journal} {\bibinfo  {journal}
  {Progress in Polymer Science}\ }\textbf {\bibinfo {volume} {30}},\ \bibinfo
  {pages} {1049} (\bibinfo {year} {2005})}\BibitemShut {NoStop}%
\bibitem [{\citenamefont {Buyukdagli}(2017)}]{buyukdagli2017like}%
  \BibitemOpen
  \bibfield  {author} {\bibinfo {author} {\bibfnamefont {S.}~\bibnamefont
  {Buyukdagli}},\ }\href@noop {} {\bibfield  {journal} {\bibinfo  {journal}
  {Physical Review E}\ }\textbf {\bibinfo {volume} {95}},\ \bibinfo {pages}
  {022502} (\bibinfo {year} {2017})}\BibitemShut {NoStop}%
\bibitem [{\citenamefont {Levin}(2005)}]{levin2005strange}%
  \BibitemOpen
  \bibfield  {author} {\bibinfo {author} {\bibfnamefont {Y.}~\bibnamefont
  {Levin}},\ }\href@noop {} {\bibfield  {journal} {\bibinfo  {journal} {Physica
  A: Statistical Mechanics and its Applications}\ }\textbf {\bibinfo {volume}
  {352}},\ \bibinfo {pages} {43} (\bibinfo {year} {2005})}\BibitemShut
  {NoStop}%
\bibitem [{\citenamefont {Luo}\ and\ \citenamefont
  {Saltzman}(2000)}]{luo2000synthetic}%
  \BibitemOpen
  \bibfield  {author} {\bibinfo {author} {\bibfnamefont {D.}~\bibnamefont
  {Luo}}\ and\ \bibinfo {author} {\bibfnamefont {W.~M.}\ \bibnamefont
  {Saltzman}},\ }\href@noop {} {\bibfield  {journal} {\bibinfo  {journal}
  {Nature Biotechnology}\ }\textbf {\bibinfo {volume} {18}},\ \bibinfo {pages}
  {33} (\bibinfo {year} {2000})}\BibitemShut {NoStop}%
\bibitem [{\citenamefont {Cakmak}\ \emph {et~al.}(2004)\citenamefont {Cakmak},
  \citenamefont {Ulukanli}, \citenamefont {Tuzcu}, \citenamefont {Karabuga},\
  and\ \citenamefont {Genctav}}]{cakmak2004synthesis}%
  \BibitemOpen
  \bibfield  {author} {\bibinfo {author} {\bibfnamefont {I.}~\bibnamefont
  {Cakmak}}, \bibinfo {author} {\bibfnamefont {Z.}~\bibnamefont {Ulukanli}},
  \bibinfo {author} {\bibfnamefont {M.}~\bibnamefont {Tuzcu}}, \bibinfo
  {author} {\bibfnamefont {S.}~\bibnamefont {Karabuga}}, \ and\ \bibinfo
  {author} {\bibfnamefont {K.}~\bibnamefont {Genctav}},\ }\href@noop {}
  {\bibfield  {journal} {\bibinfo  {journal} {European Polymer Journal}\
  }\textbf {\bibinfo {volume} {40}},\ \bibinfo {pages} {2373} (\bibinfo {year}
  {2004})}\BibitemShut {NoStop}%
\bibitem [{\citenamefont {Huang}\ \emph {et~al.}(2015)\citenamefont {Huang},
  \citenamefont {Zhong}, \citenamefont {Huang}, \citenamefont {Zhu},
  \citenamefont {Pei}, \citenamefont {Wang}, \citenamefont {Xue}, \citenamefont
  {Xie},\ and\ \citenamefont {Zhi}}]{huang2015self}%
  \BibitemOpen
  \bibfield  {author} {\bibinfo {author} {\bibfnamefont {Y.}~\bibnamefont
  {Huang}}, \bibinfo {author} {\bibfnamefont {M.}~\bibnamefont {Zhong}},
  \bibinfo {author} {\bibfnamefont {Y.}~\bibnamefont {Huang}}, \bibinfo
  {author} {\bibfnamefont {M.}~\bibnamefont {Zhu}}, \bibinfo {author}
  {\bibfnamefont {Z.}~\bibnamefont {Pei}}, \bibinfo {author} {\bibfnamefont
  {Z.}~\bibnamefont {Wang}}, \bibinfo {author} {\bibfnamefont {Q.}~\bibnamefont
  {Xue}}, \bibinfo {author} {\bibfnamefont {X.}~\bibnamefont {Xie}}, \ and\
  \bibinfo {author} {\bibfnamefont {C.}~\bibnamefont {Zhi}},\ }\href@noop {}
  {\bibfield  {journal} {\bibinfo  {journal} {Nature Communications}\ }\textbf
  {\bibinfo {volume} {6}},\ \bibinfo {pages} {10310} (\bibinfo {year}
  {2015})}\BibitemShut {NoStop}%
\bibitem [{\citenamefont {Gonz{\'a}lez}\ \emph {et~al.}(2016)\citenamefont
  {Gonz{\'a}lez}, \citenamefont {Goikolea}, \citenamefont {Barrena},\ and\
  \citenamefont {Mysyk}}]{gonzalez2016review}%
  \BibitemOpen
  \bibfield  {author} {\bibinfo {author} {\bibfnamefont {A.}~\bibnamefont
  {Gonz{\'a}lez}}, \bibinfo {author} {\bibfnamefont {E.}~\bibnamefont
  {Goikolea}}, \bibinfo {author} {\bibfnamefont {J.~A.}\ \bibnamefont
  {Barrena}}, \ and\ \bibinfo {author} {\bibfnamefont {R.}~\bibnamefont
  {Mysyk}},\ }\href@noop {} {\bibfield  {journal} {\bibinfo  {journal}
  {Renewable and Sustainable Energy Reviews}\ }\textbf {\bibinfo {volume}
  {58}},\ \bibinfo {pages} {1189} (\bibinfo {year} {2016})}\BibitemShut
  {NoStop}%
\bibitem [{\citenamefont {Vangari}\ \emph {et~al.}(2012)\citenamefont
  {Vangari}, \citenamefont {Pryor},\ and\ \citenamefont
  {Jiang}}]{vangari2012supercapacitors}%
  \BibitemOpen
  \bibfield  {author} {\bibinfo {author} {\bibfnamefont {M.}~\bibnamefont
  {Vangari}}, \bibinfo {author} {\bibfnamefont {T.}~\bibnamefont {Pryor}}, \
  and\ \bibinfo {author} {\bibfnamefont {L.}~\bibnamefont {Jiang}},\
  }\href@noop {} {\bibfield  {journal} {\bibinfo  {journal} {Journal of Energy
  Engineering}\ }\textbf {\bibinfo {volume} {139}},\ \bibinfo {pages} {72}
  (\bibinfo {year} {2012})}\BibitemShut {NoStop}%
\bibitem [{\citenamefont {Simon}\ and\ \citenamefont
  {Gogotsi}(2010)}]{simon2010materials}%
  \BibitemOpen
  \bibfield  {author} {\bibinfo {author} {\bibfnamefont {P.}~\bibnamefont
  {Simon}}\ and\ \bibinfo {author} {\bibfnamefont {Y.}~\bibnamefont
  {Gogotsi}},\ }\bibfield  {booktitle} {\emph {\bibinfo {booktitle}
  {Nanoscience And Technology: A Collection of Reviews from Nature Journals}},\
  }\href@noop {} {\ ,\ \bibinfo {pages} {320} (\bibinfo {year}
  {2010})}\BibitemShut {NoStop}%
\bibitem [{\citenamefont {Liu}\ \emph {et~al.}(2010)\citenamefont {Liu},
  \citenamefont {Yu}, \citenamefont {Neff}, \citenamefont {Zhamu},\ and\
  \citenamefont {Jang}}]{liu2010graphene}%
  \BibitemOpen
  \bibfield  {author} {\bibinfo {author} {\bibfnamefont {C.}~\bibnamefont
  {Liu}}, \bibinfo {author} {\bibfnamefont {Z.}~\bibnamefont {Yu}}, \bibinfo
  {author} {\bibfnamefont {D.}~\bibnamefont {Neff}}, \bibinfo {author}
  {\bibfnamefont {A.}~\bibnamefont {Zhamu}}, \ and\ \bibinfo {author}
  {\bibfnamefont {B.~Z.}\ \bibnamefont {Jang}},\ }\href@noop {} {\bibfield
  {journal} {\bibinfo  {journal} {Nano letters}\ }\textbf {\bibinfo {volume}
  {10}},\ \bibinfo {pages} {4863} (\bibinfo {year} {2010})}\BibitemShut
  {NoStop}%
\bibitem [{\citenamefont {Yang}\ \emph {et~al.}(2017)\citenamefont {Yang},
  \citenamefont {Kannappan}, \citenamefont {Pandian}, \citenamefont {Jang},
  \citenamefont {Lee},\ and\ \citenamefont {Lu}}]{yang2017graphene}%
  \BibitemOpen
  \bibfield  {author} {\bibinfo {author} {\bibfnamefont {H.}~\bibnamefont
  {Yang}}, \bibinfo {author} {\bibfnamefont {S.}~\bibnamefont {Kannappan}},
  \bibinfo {author} {\bibfnamefont {A.~S.}\ \bibnamefont {Pandian}}, \bibinfo
  {author} {\bibfnamefont {J.-H.}\ \bibnamefont {Jang}}, \bibinfo {author}
  {\bibfnamefont {Y.~S.}\ \bibnamefont {Lee}}, \ and\ \bibinfo {author}
  {\bibfnamefont {W.}~\bibnamefont {Lu}},\ }\href@noop {} {\bibfield  {journal}
  {\bibinfo  {journal} {Nanotechnology}\ }\textbf {\bibinfo {volume} {28}},\
  \bibinfo {pages} {445401} (\bibinfo {year} {2017})}\BibitemShut {NoStop}%
\bibitem [{\citenamefont {Jim{\'e}nez-{\'A}ngeles}\ and\ \citenamefont
  {Lozada-Cassou}(2004)}]{jimenez2004model}%
  \BibitemOpen
  \bibfield  {author} {\bibinfo {author} {\bibfnamefont {F.}~\bibnamefont
  {Jim{\'e}nez-{\'A}ngeles}}\ and\ \bibinfo {author} {\bibfnamefont
  {M.}~\bibnamefont {Lozada-Cassou}},\ }\href@noop {} {\bibfield  {journal}
  {\bibinfo  {journal} {The Journal of Physical Chemistry B}\ }\textbf
  {\bibinfo {volume} {108}},\ \bibinfo {pages} {7286} (\bibinfo {year}
  {2004})}\BibitemShut {NoStop}%
\bibitem [{\citenamefont {Colla}\ \emph {et~al.}(2016)\citenamefont {Colla},
  \citenamefont {Girotto}, \citenamefont {dos Santos},\ and\ \citenamefont
  {Levin}}]{colla2016charge}%
  \BibitemOpen
  \bibfield  {author} {\bibinfo {author} {\bibfnamefont {T.}~\bibnamefont
  {Colla}}, \bibinfo {author} {\bibfnamefont {M.}~\bibnamefont {Girotto}},
  \bibinfo {author} {\bibfnamefont {A.~P.}\ \bibnamefont {dos Santos}}, \ and\
  \bibinfo {author} {\bibfnamefont {Y.}~\bibnamefont {Levin}},\ }\href@noop {}
  {\bibfield  {journal} {\bibinfo  {journal} {The Journal of Chemical Physics}\
  }\textbf {\bibinfo {volume} {145}},\ \bibinfo {pages} {094704} (\bibinfo
  {year} {2016})}\BibitemShut {NoStop}%
\bibitem [{\citenamefont {Lee}\ \emph {et~al.}(2017)\citenamefont {Lee},
  \citenamefont {Perez-Martinez}, \citenamefont {Smith},\ and\ \citenamefont
  {Perkin}}]{perez2017scaling}%
  \BibitemOpen
  \bibfield  {author} {\bibinfo {author} {\bibfnamefont {A.~A.}\ \bibnamefont
  {Lee}}, \bibinfo {author} {\bibfnamefont {C.~S.}\ \bibnamefont
  {Perez-Martinez}}, \bibinfo {author} {\bibfnamefont {A.~M.}\ \bibnamefont
  {Smith}}, \ and\ \bibinfo {author} {\bibfnamefont {S.}~\bibnamefont
  {Perkin}},\ }\href@noop {} {\bibfield  {journal} {\bibinfo  {journal}
  {Physical Review Letters}\ }\textbf {\bibinfo {volume} {119}},\ \bibinfo
  {pages} {026002} (\bibinfo {year} {2017})}\BibitemShut {NoStop}%
\bibitem [{\citenamefont {Zwanikken}\ and\ \citenamefont {Olvera de~la
  Cruz}(2013)}]{zwanikken2013tunable}%
  \BibitemOpen
  \bibfield  {author} {\bibinfo {author} {\bibfnamefont {J.~W.}\ \bibnamefont
  {Zwanikken}}\ and\ \bibinfo {author} {\bibfnamefont {M.}~\bibnamefont {Olvera
  de~la Cruz}},\ }\href@noop {} {\bibfield  {journal} {\bibinfo  {journal}
  {Proceedings of the National Academy of Sciences}\ }\textbf {\bibinfo
  {volume} {110}},\ \bibinfo {pages} {5301} (\bibinfo {year}
  {2013})}\BibitemShut {NoStop}%
\bibitem [{\citenamefont {Antila}\ and\ \citenamefont
  {Luijten}(2018)}]{antila2018dielectric}%
  \BibitemOpen
  \bibfield  {author} {\bibinfo {author} {\bibfnamefont {H.~S.}\ \bibnamefont
  {Antila}}\ and\ \bibinfo {author} {\bibfnamefont {E.}~\bibnamefont
  {Luijten}},\ }\href@noop {} {\bibfield  {journal} {\bibinfo  {journal}
  {Physical Review Letters}\ }\textbf {\bibinfo {volume} {120}},\ \bibinfo
  {pages} {135501} (\bibinfo {year} {2018})}\BibitemShut {NoStop}%
\bibitem [{\citenamefont {Li}\ \emph {et~al.}(2016)\citenamefont {Li},
  \citenamefont {Erbaş}, \citenamefont {Zwanikken},\ and\ \citenamefont
  {Olvera de~la Cruz}}]{li2016ionic}%
  \BibitemOpen
  \bibfield  {author} {\bibinfo {author} {\bibfnamefont {H.}~\bibnamefont
  {Li}}, \bibinfo {author} {\bibfnamefont {A.}~\bibnamefont {Erbaş}},
  \bibinfo {author} {\bibfnamefont {J.}~\bibnamefont {Zwanikken}}, \ and\
  \bibinfo {author} {\bibfnamefont {M.}~\bibnamefont {Olvera de~la Cruz}},\
  }\href@noop {} {\bibfield  {journal} {\bibinfo  {journal} {Macromolecules}\
  }\textbf {\bibinfo {volume} {49}},\ \bibinfo {pages} {9239} (\bibinfo {year}
  {2016})}\BibitemShut {NoStop}%
\bibitem [{\citenamefont {Hickey}\ and\ \citenamefont
  {Holm}(2013)}]{hickey2013electrophoretic}%
  \BibitemOpen
  \bibfield  {author} {\bibinfo {author} {\bibfnamefont {O.~A.}\ \bibnamefont
  {Hickey}}\ and\ \bibinfo {author} {\bibfnamefont {C.}~\bibnamefont {Holm}},\
  }\href@noop {} {\bibfield  {journal} {\bibinfo  {journal} {The Journal of
  Chemical Physics}\ }\textbf {\bibinfo {volume} {138}},\ \bibinfo {pages}
  {194905} (\bibinfo {year} {2013})}\BibitemShut {NoStop}%
\bibitem [{\citenamefont {Qiao}\ \emph {et~al.}(2011)\citenamefont {Qiao},
  \citenamefont {Cerda},\ and\ \citenamefont {Holm}}]{qiao2011atomistic}%
  \BibitemOpen
  \bibfield  {author} {\bibinfo {author} {\bibfnamefont {B.}~\bibnamefont
  {Qiao}}, \bibinfo {author} {\bibfnamefont {J.~J.}\ \bibnamefont {Cerda}}, \
  and\ \bibinfo {author} {\bibfnamefont {C.}~\bibnamefont {Holm}},\ }\href@noop
  {} {\bibfield  {journal} {\bibinfo  {journal} {Macromolecules}\ }\textbf
  {\bibinfo {volume} {44}},\ \bibinfo {pages} {1707} (\bibinfo {year}
  {2011})}\BibitemShut {NoStop}%
\bibitem [{\citenamefont {Jing}\ \emph {et~al.}(2015)\citenamefont {Jing},
  \citenamefont {Jadhao}, \citenamefont {Zwanikken},\ and\ \citenamefont
  {Olvera de~la Cruz}}]{jing2015ionic}%
  \BibitemOpen
  \bibfield  {author} {\bibinfo {author} {\bibfnamefont {Y.}~\bibnamefont
  {Jing}}, \bibinfo {author} {\bibfnamefont {V.}~\bibnamefont {Jadhao}},
  \bibinfo {author} {\bibfnamefont {J.~W.}\ \bibnamefont {Zwanikken}}, \ and\
  \bibinfo {author} {\bibfnamefont {M.}~\bibnamefont {Olvera de~la Cruz}},\
  }\href@noop {} {\bibfield  {journal} {\bibinfo  {journal} {The Journal of
  Chemical Physics}\ }\textbf {\bibinfo {volume} {143}},\ \bibinfo {pages}
  {194508} (\bibinfo {year} {2015})}\BibitemShut {NoStop}%
\bibitem [{\citenamefont {Guerrero~García}\ and\ \citenamefont {Olvera de~la
  Cruz}(2014)}]{guerrero2014polarization}%
  \BibitemOpen
  \bibfield  {author} {\bibinfo {author} {\bibfnamefont {G.~I.}\ \bibnamefont
  {Guerrero~García}}\ and\ \bibinfo {author} {\bibfnamefont {M.}~\bibnamefont
  {Olvera de~la Cruz}},\ }\href@noop {} {\bibfield  {journal} {\bibinfo
  {journal} {The Journal of Physical Chemistry B}\ }\textbf {\bibinfo {volume}
  {118}},\ \bibinfo {pages} {8854} (\bibinfo {year} {2014})}\BibitemShut
  {NoStop}%
\bibitem [{\citenamefont {Girotto}\ \emph {et~al.}(2018)\citenamefont
  {Girotto}, \citenamefont {Malossi}, \citenamefont {dos Santos},\ and\
  \citenamefont {Levin}}]{girotto2018lattice}%
  \BibitemOpen
  \bibfield  {author} {\bibinfo {author} {\bibfnamefont {M.}~\bibnamefont
  {Girotto}}, \bibinfo {author} {\bibfnamefont {R.~M.}\ \bibnamefont
  {Malossi}}, \bibinfo {author} {\bibfnamefont {A.~P.}\ \bibnamefont {dos
  Santos}}, \ and\ \bibinfo {author} {\bibfnamefont {Y.}~\bibnamefont
  {Levin}},\ }\href@noop {} {\bibfield  {journal} {\bibinfo  {journal} {The
  Journal of Chemical Physics}\ }\textbf {\bibinfo {volume} {148}},\ \bibinfo
  {pages} {193829} (\bibinfo {year} {2018})}\BibitemShut {NoStop}%
\bibitem [{\citenamefont {dos Santos}\ \emph
  {et~al.}(2016{\natexlab{a}})\citenamefont {dos Santos}, \citenamefont
  {Girotto},\ and\ \citenamefont {Levin}}]{dos2016simulationsa}%
  \BibitemOpen
  \bibfield  {author} {\bibinfo {author} {\bibfnamefont {A.~P.}\ \bibnamefont
  {dos Santos}}, \bibinfo {author} {\bibfnamefont {M.}~\bibnamefont {Girotto}},
  \ and\ \bibinfo {author} {\bibfnamefont {Y.}~\bibnamefont {Levin}},\
  }\href@noop {} {\bibfield  {journal} {\bibinfo  {journal} {The Journal of
  Chemical Physics}\ }\textbf {\bibinfo {volume} {144}},\ \bibinfo {pages}
  {144103} (\bibinfo {year} {2016}{\natexlab{a}})}\BibitemShut {NoStop}%
\bibitem [{\citenamefont {dos Santos}\ \emph
  {et~al.}(2016{\natexlab{b}})\citenamefont {dos Santos}, \citenamefont
  {Girotto},\ and\ \citenamefont {Levin}}]{dos2016simulationsb}%
  \BibitemOpen
  \bibfield  {author} {\bibinfo {author} {\bibfnamefont {A.~P.}\ \bibnamefont
  {dos Santos}}, \bibinfo {author} {\bibfnamefont {M.}~\bibnamefont {Girotto}},
  \ and\ \bibinfo {author} {\bibfnamefont {Y.}~\bibnamefont {Levin}},\
  }\href@noop {} {\bibfield  {journal} {\bibinfo  {journal} {The Journal of
  Physical Chemistry B}\ }\textbf {\bibinfo {volume} {120}},\ \bibinfo {pages}
  {10387} (\bibinfo {year} {2016}{\natexlab{b}})}\BibitemShut {NoStop}%
\bibitem [{\citenamefont {dos Santos}\ and\ \citenamefont
  {Levin}(2015)}]{dos2015electrolytes}%
  \BibitemOpen
  \bibfield  {author} {\bibinfo {author} {\bibfnamefont {A.~P.}\ \bibnamefont
  {dos Santos}}\ and\ \bibinfo {author} {\bibfnamefont {Y.}~\bibnamefont
  {Levin}},\ }\href@noop {} {\bibfield  {journal} {\bibinfo  {journal} {The
  Journal of Chemical Physics}\ }\textbf {\bibinfo {volume} {142}},\ \bibinfo
  {pages} {194104} (\bibinfo {year} {2015})}\BibitemShut {NoStop}%
\bibitem [{\citenamefont {Messina}(2004)}]{messina2004effect}%
  \BibitemOpen
  \bibfield  {author} {\bibinfo {author} {\bibfnamefont {R.}~\bibnamefont
  {Messina}},\ }\href@noop {} {\bibfield  {journal} {\bibinfo  {journal}
  {Physical Review E}\ }\textbf {\bibinfo {volume} {70}},\ \bibinfo {pages}
  {051802} (\bibinfo {year} {2004})}\BibitemShut {NoStop}%
\bibitem [{\citenamefont {Messina}(2006)}]{messina2006erratum}%
  \BibitemOpen
  \bibfield  {author} {\bibinfo {author} {\bibfnamefont {R.}~\bibnamefont
  {Messina}},\ }\href@noop {} {\bibfield  {journal} {\bibinfo  {journal}
  {Physical Review E}\ }\textbf {\bibinfo {volume} {74}},\ \bibinfo {pages}
  {049906} (\bibinfo {year} {2006})}\BibitemShut {NoStop}%
\bibitem [{\citenamefont {Stevens}\ and\ \citenamefont
  {Kremer}(1993)}]{stevens1993structure}%
  \BibitemOpen
  \bibfield  {author} {\bibinfo {author} {\bibfnamefont {M.~J.}\ \bibnamefont
  {Stevens}}\ and\ \bibinfo {author} {\bibfnamefont {K.}~\bibnamefont
  {Kremer}},\ }\href@noop {} {\bibfield  {journal} {\bibinfo  {journal}
  {Physical Review Letters}\ }\textbf {\bibinfo {volume} {71}},\ \bibinfo
  {pages} {2228} (\bibinfo {year} {1993})}\BibitemShut {NoStop}%
\bibitem [{\citenamefont {Tyagi}\ \emph {et~al.}(2010)\citenamefont {Tyagi},
  \citenamefont {S{\"u}zen}, \citenamefont {Sega}, \citenamefont {Barbosa},
  \citenamefont {Kantorovich},\ and\ \citenamefont
  {Holm}}]{tyagi2010iterative}%
  \BibitemOpen
  \bibfield  {author} {\bibinfo {author} {\bibfnamefont {S.}~\bibnamefont
  {Tyagi}}, \bibinfo {author} {\bibfnamefont {M.}~\bibnamefont {S{\"u}zen}},
  \bibinfo {author} {\bibfnamefont {M.}~\bibnamefont {Sega}}, \bibinfo {author}
  {\bibfnamefont {M.}~\bibnamefont {Barbosa}}, \bibinfo {author} {\bibfnamefont
  {S.~S.}\ \bibnamefont {Kantorovich}}, \ and\ \bibinfo {author} {\bibfnamefont
  {C.}~\bibnamefont {Holm}},\ }\href@noop {} {\bibfield  {journal} {\bibinfo
  {journal} {The Journal of Chemical Physics}\ }\textbf {\bibinfo {volume}
  {132}},\ \bibinfo {pages} {154112} (\bibinfo {year} {2010})}\BibitemShut
  {NoStop}%
\bibitem [{\citenamefont {Frenkel}\ and\ \citenamefont
  {Smit}(2001)}]{frenkel2001understanding}%
  \BibitemOpen
  \bibfield  {author} {\bibinfo {author} {\bibfnamefont {D.}~\bibnamefont
  {Frenkel}}\ and\ \bibinfo {author} {\bibfnamefont {B.}~\bibnamefont {Smit}},\
  }\href@noop {} {\emph {\bibinfo {title} {Understanding molecular simulation:
  from algorithms to applications}}},\ Vol.~\bibinfo {volume} {1}\ (\bibinfo
  {publisher} {Elsevier},\ \bibinfo {year} {2001})\BibitemShut {NoStop}%
\bibitem [{\citenamefont {Nguyen}\ \emph {et~al.}(2019)\citenamefont {Nguyen},
  \citenamefont {Li}, \citenamefont {Bagchi}, \citenamefont {Solis},\ and\
  \citenamefont {Olvera de~la Cruz}}]{nguyen2019incorporating}%
  \BibitemOpen
  \bibfield  {author} {\bibinfo {author} {\bibfnamefont {T.~D.}\ \bibnamefont
  {Nguyen}}, \bibinfo {author} {\bibfnamefont {H.}~\bibnamefont {Li}}, \bibinfo
  {author} {\bibfnamefont {D.}~\bibnamefont {Bagchi}}, \bibinfo {author}
  {\bibfnamefont {F.~J.}\ \bibnamefont {Solis}}, \ and\ \bibinfo {author}
  {\bibfnamefont {M.}~\bibnamefont {Olvera de~la Cruz}},\ }\href@noop {}
  {\bibfield  {journal} {\bibinfo  {journal} {Computer Physics Communications}\
  } (\bibinfo {year} {2019})}\BibitemShut {NoStop}%
\bibitem [{\citenamefont {Guerrero-Garc{\'\i}a}\ \emph
  {et~al.}(2010{\natexlab{a}})\citenamefont {Guerrero-Garc{\'\i}a},
  \citenamefont {Gonz{\'a}lez-Tovar}, \citenamefont {Ch{\'a}vez-P{\'a}ez},\
  and\ \citenamefont {Lozada-Cassou}}]{guerrero2010overcharging}%
  \BibitemOpen
  \bibfield  {author} {\bibinfo {author} {\bibfnamefont {G.~I.}\ \bibnamefont
  {Guerrero-Garc{\'\i}a}}, \bibinfo {author} {\bibfnamefont {E.}~\bibnamefont
  {Gonz{\'a}lez-Tovar}}, \bibinfo {author} {\bibfnamefont {M.}~\bibnamefont
  {Ch{\'a}vez-P{\'a}ez}}, \ and\ \bibinfo {author} {\bibfnamefont
  {M.}~\bibnamefont {Lozada-Cassou}},\ }\href@noop {} {\bibfield  {journal}
  {\bibinfo  {journal} {The Journal of Chemical Physics}\ }\textbf {\bibinfo
  {volume} {132}},\ \bibinfo {pages} {054903} (\bibinfo {year}
  {2010}{\natexlab{a}})}\BibitemShut {NoStop}%
\bibitem [{\citenamefont {Guerrero-Garc{\'\i}a}\ \emph
  {et~al.}(2010{\natexlab{b}})\citenamefont {Guerrero-Garc{\'\i}a},
  \citenamefont {Gonz{\'a}lez-Tovar},\ and\ \citenamefont {Olvera de~la
  Cruz}}]{guerrero2010effects}%
  \BibitemOpen
  \bibfield  {author} {\bibinfo {author} {\bibfnamefont {G.~I.}\ \bibnamefont
  {Guerrero-Garc{\'\i}a}}, \bibinfo {author} {\bibfnamefont {E.}~\bibnamefont
  {Gonz{\'a}lez-Tovar}}, \ and\ \bibinfo {author} {\bibfnamefont
  {M.}~\bibnamefont {Olvera de~la Cruz}},\ }\href@noop {} {\bibfield  {journal}
  {\bibinfo  {journal} {Soft Matter}\ }\textbf {\bibinfo {volume} {6}},\
  \bibinfo {pages} {2056} (\bibinfo {year} {2010}{\natexlab{b}})}\BibitemShut
  {NoStop}%
\bibitem [{\citenamefont {Messina}\ \emph {et~al.}(2002)\citenamefont
  {Messina}, \citenamefont {Gonz{\'a}lez-Tovar}, \citenamefont
  {Lozada-Cassou},\ and\ \citenamefont {Holm}}]{messina2002overcharging}%
  \BibitemOpen
  \bibfield  {author} {\bibinfo {author} {\bibfnamefont {R.}~\bibnamefont
  {Messina}}, \bibinfo {author} {\bibfnamefont {E.}~\bibnamefont
  {Gonz{\'a}lez-Tovar}}, \bibinfo {author} {\bibfnamefont {M.}~\bibnamefont
  {Lozada-Cassou}}, \ and\ \bibinfo {author} {\bibfnamefont {C.}~\bibnamefont
  {Holm}},\ }\href@noop {} {\bibfield  {journal} {\bibinfo  {journal} {EPL
  (Europhysics Letters)}\ }\textbf {\bibinfo {volume} {60}},\ \bibinfo {pages}
  {383} (\bibinfo {year} {2002})}\BibitemShut {NoStop}%
\bibitem [{\citenamefont {Veerman}\ and\ \citenamefont
  {Frenkel}(1992)}]{veerman1992phase}%
  \BibitemOpen
  \bibfield  {author} {\bibinfo {author} {\bibfnamefont {J.}~\bibnamefont
  {Veerman}}\ and\ \bibinfo {author} {\bibfnamefont {D.}~\bibnamefont
  {Frenkel}},\ }\href@noop {} {\bibfield  {journal} {\bibinfo  {journal}
  {Physical Review A}\ }\textbf {\bibinfo {volume} {45}},\ \bibinfo {pages}
  {5632} (\bibinfo {year} {1992})}\BibitemShut {NoStop}%
\bibitem [{\citenamefont {Partenskii}\ \emph {et~al.}(1996)\citenamefont
  {Partenskii}, \citenamefont {Dorman},\ and\ \citenamefont
  {Jordan}}]{partenskii1996question}%
  \BibitemOpen
  \bibfield  {author} {\bibinfo {author} {\bibfnamefont {M.~B.}\ \bibnamefont
  {Partenskii}}, \bibinfo {author} {\bibfnamefont {V.}~\bibnamefont {Dorman}},
  \ and\ \bibinfo {author} {\bibfnamefont {P.~C.}\ \bibnamefont {Jordan}},\
  }\href@noop {} {\bibfield  {journal} {\bibinfo  {journal} {International
  Reviews in Physical Chemistry}\ }\textbf {\bibinfo {volume} {15}},\ \bibinfo
  {pages} {153} (\bibinfo {year} {1996})}\BibitemShut {NoStop}%
\bibitem [{\citenamefont {Partenskii}\ and\ \citenamefont
  {Jordan}(2008)}]{partenskii2008limitations}%
  \BibitemOpen
  \bibfield  {author} {\bibinfo {author} {\bibfnamefont {M.~B.}\ \bibnamefont
  {Partenskii}}\ and\ \bibinfo {author} {\bibfnamefont {P.~C.}\ \bibnamefont
  {Jordan}},\ }\href@noop {} {\bibfield  {journal} {\bibinfo  {journal}
  {Physical Review E}\ }\textbf {\bibinfo {volume} {77}},\ \bibinfo {pages}
  {061117} (\bibinfo {year} {2008})}\BibitemShut {NoStop}%
\bibitem [{\citenamefont {Partenskii}\ and\ \citenamefont
  {Jordan}(2009)}]{partenskii2009squishy}%
  \BibitemOpen
  \bibfield  {author} {\bibinfo {author} {\bibfnamefont {M.~B.}\ \bibnamefont
  {Partenskii}}\ and\ \bibinfo {author} {\bibfnamefont {P.~C.}\ \bibnamefont
  {Jordan}},\ }\href@noop {} {\bibfield  {journal} {\bibinfo  {journal}
  {Physical Review E}\ }\textbf {\bibinfo {volume} {80}},\ \bibinfo {pages}
  {011112} (\bibinfo {year} {2009})}\BibitemShut {NoStop}%
\bibitem [{\citenamefont {Hatlo}\ and\ \citenamefont
  {Lue}(2008)}]{hatlo2008role}%
  \BibitemOpen
  \bibfield  {author} {\bibinfo {author} {\bibfnamefont {M.~M.}\ \bibnamefont
  {Hatlo}}\ and\ \bibinfo {author} {\bibfnamefont {L.}~\bibnamefont {Lue}},\
  }\href@noop {} {\bibfield  {journal} {\bibinfo  {journal} {Soft Matter}\
  }\textbf {\bibinfo {volume} {4}},\ \bibinfo {pages} {1582} (\bibinfo {year}
  {2008})}\BibitemShut {NoStop}%
\end{thebibliography}
\end{document}